\documentclass[a4paper,12pt]{article}
\usepackage{graphicx,color}
\usepackage{amssymb}
\usepackage{amsmath}

\usepackage{cite}

\usepackage{authblk}

\def\noi{\noindent}
\def\e{{\,\rm e}}

\def\const{{\rm const}}
\def\eps{\varepsilon}
\def\ep{\epsilon}

\def\then{\ \Rightarrow\ }

\def\nqq{\hspace*{-2em}}
\def\nhq{\hspace*{-0.5em}}

\def\al{&\nhq}
\def\lal{&&\nqq {}}
\def\eq{Eq.\,}
\def\eqs{Eqs.\,}
\def\beq{\begin{equation}}
\def\eeq{\end{equation}}
\def\bear{\begin{eqnarray}}
\def\bearr{\begin{eqnarray} \lal}
\def\ear{\end{eqnarray}}
\def\earn{\nonumber \end{eqnarray}}

\def\yyy{\\[5pt] \lal }
\def\eql{\al =\al}

\def\eqn#1{\eq\eqref{#1}}
\def\rf{\eqref}

\def\mN{_{\mu}^{\nu}}

\def\N{{\mathbb N}}
\def\R{{\mathbb R}}

\def\asflat{asymptotically flat}

\def\ssph{static, spherically symmetric}
\def\bh{black hole}
\def\bhs{black holes}

\def\ssph{static, spherically symmetric}

\def\intco{integration constant}
\def\elmag{electromagnetic}
\def\asflat{asymptotically flat}

\begin{document}
\begin{titlepage}


\title{Static, spherically symmetric solutions with a scalar field in Rastall gravity}

\author[1,3,4]{\small\bf K.A. Bronnikov\thanks{kb20@yandex.ru}}
\author[2,4]{\small\bf J.C. Fabris\thanks{fabris@pq.cnpq.br}}
\author[2]{\small\bf O.F. Piattella\thanks{oliver.piattella@pq.cnpq.br}}
\author[2]{\small\bf E.C. Santos\thanks{edison\_cesar@hotmail.com}}

\affil[1]{\small\it VNIIMS, Ozyornaya ul. 46, Moscow 119361, Russia}
\affil[2]{\small\it Departamento de F\'isica, UFES, Avenida Fernando Ferrari, 514, 
	CEP 29075-910, Vit\'oria, ES, Brazil.}
\affil[3]{\small\it Institute of Gravitation and Cosmology, PFUR, ul. Miklukho-Maklaya 6, 
	Moscow 117198, Russia}
\affil[4]{\small\it  National Research Nuclear University - MEPhI, Kashirskoe sh. 31, 
	Moscow 115409, Russia}
\maketitle
\vspace{-5mm}

\begin{abstract}
     Rastall's theory belongs to the class of non-conservative theories of gravity. In vacuum, the only 
     non-trivial static, spherically symmetric solution is the Schwarzschild one, except for a very 
     special case. When a canonical scalar field is coupled to the gravity sector in this theory, 
     new exact solutions appear for some values of the Rastall parameter $a$.
     Some of these solutions describe the same space-time geometry as the recently found 
     solutions in the $k$-essence theory with a power function for the kinetic term of the scalar field. 
     There is a large class of solutions (in particular, those describing wormholes and regular
     black holes) whose geometry coincides with that of solutions of GR coupled to scalar fields 
     with nontrivial self-interaction potentials; the form of these potentials, however, depends on 
     the Rastall parameter $a$. We also note that all solutions of GR with a zero trace of the
     energy-momentum tensor, including black-hole and wormhole ones, may be re-interpreted as 
     solutions of Rastall's theory.
\end{abstract}

\mbox{Pacs:\,98.80.-k, 04.50.Kd}

\end{titlepage}

\section{Introduction}

  The coupling of a scalar field to the gravitational action in a geometric context has been intensively
  studied since the emergence of the General Relativity (GR) theory. A relativistic scalar theory of 
  gravity was proposed just before the formulation of GR \cite{nord}. Much later, the 
  Jordan-Brans-Dicke theory \cite{bd} was proposed in order to incorporate the idea of a variable
  gravitational coupling \cite{dirac} as an extension of GR, using a scalar field non-minimally coupled 
  to the gravitational sector. Recently, the Horndeski class of theories \cite{horn}, which is the 
  most general combination of a scalar field and its derivative leading to second-order differential
  equations, called a lot of attention, together with the Galileon theories which are, in a sense, a 
  subclass of the former. The very fashionable $f(R)$ theory, with a nonlinear generalization of 
  the Einstein-Hilbert Lagrangian \cite{f(R)}, can be recast in the form of a scalar-tensor theory. 
  This very brief list of the use of scalar fields in the gravitational context does not exhaust the 
  huge literature on this subject.

  Another possibility that has been exploited since some years is a generalization of the kinetic 
  expression for the scalar field which is minimally coupled to the Einstein-Hilbert Lagrangian, 
  the so-called $k$-essence theories \cite{k1}. Originally, this proposal was applied as an alternative 
  to the usual inflationary models based on a self-interacting scalar field \cite{k1,k2,k3}. In a recent 
  paper, vacuum static spherically symmetric solutions have been obtained for the $k$-essence 
  theories \cite{denis}. Some new configurations were obtained, including structures 
  where an event horizon is present, but the interpretation as a black hole still seems impossible,
  mainly due to the impossibility to define a distant region from the horizon. A {\it no-go} theorem 
  has been proved, showing that solutions with a regular horizon (including the behavior of the 
 energy-momentum tensor) can be only of the {\it cold black hole} type \cite{kirill}.

  On the other hand, one more possible generalization of GR is to relax the constraint given by the
  conservation laws expressed by the zero divergence of the energy-momentum tensor. One proposal 
  in this sense is Rastall's theory \cite{rastall}. In this theory, the divergence of the energy-momentum
  tensor is linked with the gradient of the Ricci scalar, and in this way Rastall's theory may be seen as 
  a phenomenological implementation of some quantum effects in a curved background.
  At the cosmological level, Rastall's theory gives very interesting results. For example, the evolution of 
  small Dark Matter fluctuations is identical to that in the $\Lambda$CDM model, but in Rastall's theory
  Dark Energy is able to cluster. This potentially provides an evolution of Dark Matter inhomogeneities 
  in a nonlinear regime which is different from the one in the standard CDM model \cite{cosmo}. The 
  whole success of the $\Lambda$CDM model is reproduced at the background and linear perturbation
  levels, while some new effects are expected at the nonlinear level where the $\Lambda$CDM model
  faces some difficulties \cite{cusp,satt,satellites}. Moreover, the only vacuum static spherically
  symmetric solution is the Schwarzschild one, except for a very particular value of the arbitrary 
  constant characterizing this theory \cite{adriano}. 

  In Ref. \cite{oliver}, the use of the canonical energy-momentum tensor of a scalar field in Rastall's
  theory has been considered in a cosmological context and one interesting result has been obtained: 
  at a perturbative level, 
  such a structure is consistent only in presence of matter. One interesting aspect of this analysis 
  is that the resulting coupling of the scalar field with gravity leads to equations very similar to 
  those in some classes of Galileon theories.

  In the present paper, we consider the problem of finding static, spherically symmetric spaces in 
  Rastall's theory in the presence of a free or self-interacting scalar field. We will verify that some exact
  solutions can be obtained, and some of them reproduce completely, in the geometric sense, the 
  solutions recently found in the $k$-essence theory. These geometries, including the emerging global  
  causal structure, were described in detail in \cite{denis}.
  However, the behavior of the scalar field 
  is different in the two different theories, therefore, a no-go theorem proved for the $k$-essence 
  theory acquires new features in the Rastall-scalar field system. Other properties of the scalar field 
  may also have consequences for the stability of these static, spherically symmetric configurations. 
  Furthermore, considering a self-interacting scalar field, it is shown that the regular black hole
  structures, found in Ref. \cite{prl} for a phantom field, occur in this Rastall-scalar field system as 
  well, the only difference being in the form of the scalar field potential. Again, such a distinction from
  the solutions obtained in Ref.~\cite{prl} may affect their stability. 
  In the GR context, a corresponding problem was studied in \cite{BKZ}, and a family of linearly stable solutions was revealed. 

  This paper is organized as follows. In the next section we present the Rastall-scalar field equations
  and write down the equations for static, spherically symmetric configurations. In Section 3, 
  some general features of the system with and without a potential  are determined. In Section 4, 
  special exact solutions are obtained for a massless scalar field. In Section 5, regular solutions with a 
  nonzero scalar field potential are obtained by using a particular form of the radial function. 
  In Section 6 we present our conclusions.

\section{Field equations}

  Rastall's theory of gravity is characterized by the following equations \cite{rastall}:
\bearr                                                                        \label{EE-Ra}
	R_{\mu\nu} - \frac{\lambda}{2}g_{\mu\nu}R = \kappa T_{\mu\nu},
\\ \lal
	{T^{\mu\nu}}_{;\mu} = \frac{1 - \lambda}{2\kappa}R^{,\nu} 
        =  \frac{1-\lambda}{2(1-2\lambda)} T^{,\nu},
\ear
  where $\lambda$ is a free parameter. When $\lambda = 1$, GR is recovered \footnote
           {The connection between $\kappa$ and the Newtonian gravitational coupling is given by 
           \cite{rastall} $\kappa =  \dfrac{(2\lambda - 1)}{(3\lambda - 2)}\,8\pi G = \dfrac{8\pi G}{a}$. 
           The special cases $\lambda  = 1/2$ and $\lambda  = 2/3$, corresponding to infinite or
           zero values of $a$, respectively, will be discussed below.}
   From now on, we fix  $\kappa = 1$. These equations can be rewritten as \cite{al1,al2},
\bearr
	R_{\mu\nu} - \frac{1}{2}g_{\mu\nu}R =  T_{\mu\nu} - \frac{a - 1}{2}g_{\mu\nu}T,
\\ \lal
          {T^{\mu\nu}}_{;\mu} =   \frac{a - 1}{2(2a - 3)}R^{,\nu} = \frac{a-1}{2} T^{,\nu},\qquad 
                  a: = \frac{3\lambda - 2}{2\lambda - 1}.
\ear
  In this parametrization, GR is recovered if $a = 1$. It must be stressed that a Lagrangian formulation 
  for the above equations is still lacking. In Ref.~\cite{Smalley}, an action which is not an invariant 
  scalar was proposed. A possibility is that the equations of motion of Rastall's theory could be 
  derived from a variational principle in a Weyl-Cartan theory of gravity, i.e., when dropping the 
  metricity condition for the connection and allowing for the existence of torsion, breaking the usual
  diffeomorphism invariance, which is a basis for the conservation laws of GR. However, there is still 
  no clear Lagrangian formulation for Rastall's theory. There are works in progress, trying to obtain a
  consistent Lagrangian formulation for this class of nonconservative theories of gravity.

  In what follows, we use the canonical energy-momentum tensor of a self-interacting scalar field:
\begin{eqnarray}
T_{\mu\nu} = \epsilon\biggr\{\phi_{;\mu}\phi_{;\nu} - \frac{1}{2}g_{\mu\nu}\phi^{;\rho}\phi_{;\rho}\biggl\} + g_{\mu\nu}V(\phi).
\end{eqnarray}
  The parameter $\epsilon$ can take the values $\pm 1$, indicating an ordinary ($+1$) or phantom 
  ($-1$) scalar field. Using this expression for the energy-momentum tensor, the field equations read
\begin{eqnarray}
         R_{\mu\nu} - \frac{1}{2}g_{\mu\nu}R &=& \epsilon\biggr\{\phi_{;\mu}\phi_{;\nu} 
                 - \frac{2 - a}{2}g_{\mu\nu}\phi^{;\rho}\phi_{;\rho}\biggl\}
\nonumber\\
                                   &&\ \ \  + (3 - 2a)g_{\mu\nu}V(\phi),\\
                   \Box\phi + (a - 1)\frac{\phi^{;\rho}\phi^{;\sigma}\phi_{;\rho;\sigma}}
                     {\phi^{;\alpha}\phi_{;\alpha}} &=& - \epsilon(3 - 2a)V_\phi,
\end{eqnarray}
  where the subscript $\phi$ denotes $d/d\phi$.

  Since we are interested in static, spherically symmetric solutions, we consider the general metric
  representing this symmetry,
\begin{eqnarray}                                                 \label{ds}
	ds^2 = e^{2\gamma(r)}dt^2 - e^{2\alpha(r)}dr^2 - e^{2\beta(r)}d\Omega^2,
\end{eqnarray}
  where $\alpha$, $\beta$ and $\gamma$ are generic functions of the radial coordinate.

  With this metric, the field equations reduce to the following set of non-linear differential equations:
\begin{eqnarray}                   \label{eq1}
- 2\beta'' - 3{\beta'}^2 + 2\alpha'\beta' + e^{2(\alpha - \beta)} \eql 
                 \epsilon\frac{2 {-} a}{2}{\phi'}^2 + (3 {-} 2a)Ve^{2\alpha},
\\                \label{eq2}
         2\gamma'\beta' + {\beta'}^2 - e^{2(\alpha - \beta)} \eql
                                \epsilon\frac{a}{2}{\phi'}^2 - (3 - 2a)e^{2\alpha}V,
\\                \label{eq3}
              \beta'' + \gamma'' + \beta'(\gamma' + \beta' - \alpha') + \gamma'(\gamma' - \alpha') \eql 
                           - \epsilon\frac{2 - a}{2}{\phi'}^2 
\nonumber\\
                       &&\ \ \  - (3 - 2a)e^{2\alpha}V,
\\                \label{eq4}
             a\phi'' + (\gamma' + 2\beta' - a\alpha')\phi' \eql    (3 - 2a)e^{2\alpha}V_\phi. 
\end{eqnarray}
  Let us remark that there is still a freedom in choosing the radial coordinate $u$, which can be done 
  by fixing one condition on the metric functions.

  The set of equations (\ref{eq1})--(\ref{eq4}) depends on the specification of the potential
  $V(\phi)$.  It is a very complex set of equations, which can be solved exactly only for 
  some particular choices of $V(\phi)$; moreover, even with $V(\phi) = \const $
  it can be integrated only for some values of the parameter $a$. In what follows we will first 
  reveal some general features of the possible solutions of the system, 
  and then we will obtain some exact solutions.

\section{Some general features}
\subsection{Global structure}

  In Ref. \cite{denis} it has been shown that in GR with a $k$-essence scalar field without a potential,
  a horizon with finite area implies a singular behavior of the scalar field.  
  If we require regularity of the k-essence scalar field on the horizon, only cold 
  black holes with an infinite area of the horizon \cite{kirill} are possible.
  As we shall see, in the present case some general observations can also be made, and, in particular,
  horizons with a finite area can occur in a certain range of the Rastall parameter $a$.

  To consider horizons, it is helpful to use the so-called quasiglobal coordinate characterized by 
  the gauge condition $\alpha + \gamma =0$ \cite{BR-book}. In this case, denoting the coordinate
  by $x$ (to be distinguished from the general case where it is denoted by $u$) and introducing the 
  notations
\beq                                                                                  \label{def_A,r}
               e^{2\gamma} = A(x), \qquad    e^\beta = r(x), 
\eeq
   the metric takes the form
\beq                                                                                  \label{ds-q}
              ds^2 = A(x) dt^2 - dx^2/A(x) - r(x)^2 d\Omega^2, 
\eeq
  and the field equations are rewritten as follows:
\begin{eqnarray}                  \label{eb1}
         2\frac{r''}{r} + \frac{A'}{A}\frac{r'}{r} + \biggr(\frac{r'}{r}\biggl)^2  - \frac{1}{r^2 A} 
                          &=& - \epsilon\frac{2 - a}{2}{\phi'}^2 - (3 - 2a)\frac{V}{A},
\\                                        \label{eb2}
          \frac{A'}{A}\frac{r'}{r} + \biggr(\frac{r'}{r}\biggl)^2 - \frac{1}{r^2A} 
                  &=& \epsilon\frac{a}{2}{\phi'}^2 - (3 - 2a)\frac{V}{A},
\\                                       \label{eb3}
               \frac{r''}{r} + \frac{A'}{A}\frac{r'}{r} + \frac{1}{2}\frac{A''}{A} 
                               &=& - \epsilon\frac{2 - a}{2}{\phi'}^2 - (3 - 2a)\frac{V}{A},
\\                                      \label{eb4}
                   a\phi'' + \biggr\{\frac{(1 + a)}{2}\frac{A'}{A} + 2\frac{r'}{r}\biggl\}\phi' 
                               &=& (3 - 2a) \frac{V_\phi}{A}.
\end{eqnarray}
  where the prime denotes $d/dx$. It is important that the differences between \eqs \rf{eb1} and \rf{eb3}  and between \rf{eb1} and \rf{eb2}
  depend neither on the parameter $a$ nor on the potential $V(\phi)$:
\bearr                        \label{0-1}
                    2\frac{r''}{r} = - \epsilon{\phi'}^2,
\yyy                           \label{0-2}
		  A'' r^2 - A (r^2)'' = -2
\ear
  It was \eqn{0-2} that has allowed for proving the so-called {\it Global Structure Theorem}
  \cite{k-glob, BR-book} for scalar field configurations in GR. The theorem restricts possible
  black hole structures irrespective of the form of the potential. It claims that {\it if there is a 
  static region of space-time (such that $A(x) > 0$), then (i) only simple horizons are possible,
  and (ii) the number of horizon cannot be larger than 2. }

  Let us recall here that a Killing horizon, which may be an event horizon of a black hole, 
  corresponds in the metric \rf{ds} to a regular zero of the function $A(x)$, and it is simple if
  this zero is simple (that is, $A'(x) \ne 0$ where $A(x) =0$).

  We thus conclude that this theorem holds in the system under consideration for any value of
  the Rastall parameter $a$ and any choice of the potential $V(\phi)$.    

\subsection{Possible horizons with a massless field}
\label{3.2}

  Our {\it second observation} concerns the case where $V = \const$, which, if nonzero, can be
  interpreted as a cosmological constant.   . 

  In this case, the scalar equation  (\ref{eb4}) has the first integral
\begin{eqnarray}                                       \label{int-phi}
                 \phi' =  C\, A^{-(1 + a)/(2a)}\,r^{-2/a}, \qquad C = \const.
\end{eqnarray}
  A possible horizon is characterized by $A = 0$. If it occurs at a finite radius, 
  then $\phi' \to \infty$ if $a > 0$ or $a < - 1$. As to the energy-momentum tensor, 
  its physical components (the scalar field energy density and pressure) depend on $A\phi'^2$. 
This quantity is proportional to $A^{-1/a}$ and blows up at vanishing $A$ if $a > 0$.
  Thus the range of $a$ where both $\phi$ and $T\mN$ are finite at a possible horizon is only 
  $-1 < a < 0$.
 
  For $a$ outside this interval, we obtain an analog of the no-go theorem found in Ref. \cite{denis} 
  for systems with k-essence: a horizon with finite area can occurs only with singular values of 
  $\phi$ and/or $T\mN$. However, in the range $0 > a > - 1$, it is possible to have horizons with 
  finite radius, at which $T\mN \to 0$.
 
  Our reasoning did not work for the special cases $a =0$ and $a =-1$ which will be considered 
  separately below. As to the case $a=1$ (GR), the corresponding solution for a massless scalar field 
  (see, e.g., \cite{fish, k-73}) does not contain a horizon.

  Now let us turn to \eqn{0-1}: it shows that $r'' \to \infty$ if either $a>0$ or $a < -1$, and it turns to
  zero for $-1 < a < 0$. In the first case, this points at a possible cold black hole configuration if 
  $\epsilon = - 1$ and at a singular center if $\epsilon = + 1$. In the second case, the vanishing of 
  $r''$ on the horizon seems to be consistent with a regular horizon of finite radius.

  There is, however,  one more restriction: the metric, to be extendible beyond the horizon, should 
  be analytic in the quasiglobal coordinate $x$ \cite{BR-book}, therefore, if $r$ is finite, we should 
  have   $r'' \sim x^n$, $n \in \N$, or since (according to the Global Structure theorem) $A\sim x$,
  we should have $r'' \sim A^n$. According to \rf{0-1} and \rf{int-phi}, this happens if and only if
  $a = -1/(n+1)$, which belongs to the admissible range. Thus {\it a regular horizon of finite area 
  can only occur at discrete values of $a$ belonging to the sequence $\{a_n\} = \{-1/(n+1)\}$. } 

\subsection {Asymptotic behavior}

  Another problem is connected with the asymptotic behavior of  possible solutions for a scalar field 
  with zero or constant potential, for which we have the relations \eqref{0-1} and \eqref{int-phi},
  at large radii. Indeed, in an asymptotically flat space-time with the metric \eqref{ds-qua} we can
  assume the following analytic behavior of the metric functions as $x \to \infty$:
\begin{eqnarray}
		A \to 1, \qquad r(x) = x + r_0 + r_k x^{-k} + o(x^{-k}),
\end{eqnarray}
  where $r_0, r_k$ are constants and $k$ is some positive integer. This means that the l.h.s. of 
  Eq. \eqref{0-1} behaves at large $x$ in the following way:
\[
              \frac{r''}{r} = \frac{k(k+1) r_k}{x^{k+3}} (1+ o(1)).
\]
  On the other hand, from \eqref{int-phi} it follows for the r.h.s. of \eqref{0-1} that
\[
		\phi'{}^2 = C^2 x^{-4/a} (1+ o(1)).
\]
  Comparing the exponents of $x$ in these expressions, we obtain for the input constant $a$ 
  of Rastall's theory
\begin{eqnarray}
		a = \frac {4}{k+3}, \qquad    k \in {\mathbb N}.
\end{eqnarray}
  Only with such values of $a$ a scalar-vacuum solution with a massless scalar field can have an
  analytical flat asymptotic behavior. Thus, for $k = 1,2,3,4,...$ one should have $a = 1, 4/5, 2/3, 4/7,...$,
  respectively; for all other $a$ at a possible flat asymptotic the function $r(x)$ will not 
  behave analytically.

  We can conclude that if there is a regular horizon in a particular solution, this solution cannot have 
  an analytic flat asymptotic behavior.

  Even more than that: as we saw, regular horizons require $a < 0$, but in this case, if $A\to \const>0$ 
  at large $x$, then $\phi'\to \infty$, which is incompatible with a flat asymptotic. Thus we have a no-go theorem: {\it in Rastall's theory with a massless
  scalar field there cannot be a \ssph, \asflat\ black hole}. 
   
  Let us, however, recall that some special cases ($a=0, -1$) were so far not included into consideration.

\section{Special solutions}
\subsection{Special solution I: $a = -1$}

   To find the first special solution, we fix $V = 0$ and use the quasiglobal coordinate $x$, so that
   again, according to \rf{def_A,r}, $e^{2\gamma} = e^{-2\alpha} = A(x)$, and $e^{2\beta} = r^2(x)$.
   In this case, the equations of motion have the form \rf{eb1}--\rf{eb4} with $V =0$, and their
   two consequences \rf{0-1} and \rf{0-2} hold, as before.

   Also, as before, we have the first integral of the scalar equation \rf{int-phi}.
   If we choose $a = - 1$, the $A$ dependence drops out, and this relation reduces to
\begin{eqnarray}
                      \phi' = C r^2, \qquad C = \const.
\end{eqnarray}
   So, till the end of this subsection,  we focus on the case $a = -1$.

   Now, \eqn{0-1} leads to the relation
\begin{eqnarray}
                   r'' = - \frac{\epsilon}{2}C^2 r^5.
\end{eqnarray}
    Its first integral is
\begin{eqnarray}
              {r'}^2 = - \frac{\epsilon}{6}C^2r^6 + K_1,
\end{eqnarray}
  where $K_1$ is an integration constant. This equation can be further integrated, but with
  $K_1 \ne 0$ it leads to rather complex expressions with elliptic integrals, which we will not 
  consider here. Assuming $K_1 = 0$, we easily obtain 
\begin{eqnarray}
                  r = \biggr(\frac{3}{2C^2}\biggl)^{1/4} \frac{1}{\sqrt{x}}.
\end{eqnarray}
  Hence,
\begin{eqnarray}
           \phi' = \sqrt{\frac 32}\frac{1}{x} \quad \then \quad \phi = \sqrt{\frac 32}\ln x + \phi_0.
            \quad \phi_0 = \const.
\end{eqnarray}

  Now, using \eq (\ref{eb2}) and the solutions already found for $r$ and $\phi$, we have the 
  following equation:
\begin{eqnarray}
                  A' + \frac{A}{x} = - \sqrt{\frac{8 C^2}{3}} x^2.
\end{eqnarray}
  Its solution is
\begin{eqnarray}
\label{ss1}
            A = \frac{K}{x} - \frac{C}{\sqrt{6}} x^3,
\end{eqnarray}
  where $K$ is one more integration constant. Thus the metric can be written as
\begin{eqnarray}                         \label{mp1}
               ds^2 = A(x) dt^2 - A(x)^{-1}dx^2 - \sqrt{\frac{3}{2C^2}}\frac{1}{x}d\Omega^2,
\end{eqnarray}
   with $A(x)$ given by (\ref{ss1}).

  The geometry described by ({\ref{mp1}) is the same as the one found in Ref. \cite{denis}
  and presented there in some detail. The spatial infinities are singular, being separated by a horizon. 
 
  There is, however, a difference with respect to Ref. \cite{denis}: the behavior of the scalar field, 
  which in the present case is defined in the whole range $- \infty < \phi < + \infty$, passing zero, 
  while in the $k$-essence case it is always negative, never reaching zero and diverging at the 
  extremes of the range of $x$. These different features may affect the stability of the solution.

  This first special solution described previously has been obtained for $V = 0$ and $a = - 1$. 
  It is on the borderline of the two possibilities sketched in Subsection \ref{3.2}. 
If we try to describe small deviations from this borderline, then, 
  writing $a = - 1 + \eps$, performing an expansion in terms of $\eps > 0$ and considering the 
  leading term, we can verify that  $\phi' \to \const$ with a finite value. Moreover, the horizon 
  has a finite area. Hence, in opposition to the result of Ref. \cite{denis}, the present special solution 
  corresponds to a regular horizon (and even with a finite scalar field) with finite area.

\subsection{Special solution II: $a=0$}

  This case implies $\lambda = 2/3$. In principle, it corresponds to an anomalous relation 
  between $\kappa$ and the Newtonian gravitational coupling $G$ ($\kappa$ must be infinite for 
  obtaining a finite value of $G$). However, the field equations are well behaved, but the Newtonian 
  limit must be studied separately for this particular value of $\lambda$, since in the Newtonian
  approximation we obtain a Laplace equation, i.e., no matter source. This case has some 
  similarities with the other critical point of $a$, corresponding to $\lambda = 1/2$, with which 
  everything is well-behaved except for the fact that the field equations only admit coupling to 
  conformal matter with $T=0$, see \cite{adriano}.

  Let us now consider the coordinate condition $\alpha = \gamma + 2\beta$, called the {\it harmonic
  coordinate condition} \cite{k-73}. To distinguish it from the previous case, we will denote the radial
  coordinate in this gauge by $u$.

  Fixing $a = 0$ (till the end of this subsection) and considering a constant potential $V = \Lambda/3$, 
  \eq (\ref{eq4}) implies, under this coordinate condition,
\begin{eqnarray}
                    \phi'\alpha' = 0.
\end{eqnarray}
  Hence, there are two options: $\phi' = 0$, which leads to the Schwarzschild solution of GR; or 
  $\alpha' = 0$. On the other hand, this condition implies
  $\gamma + 2\beta = 0$.  The equations of motion reduce to
\begin{eqnarray}                  \label{ec1}
                  - 2\beta'' - 3{\beta'}^2 + e^{-2\beta} &=& \epsilon{\phi'}^2 + \Lambda,
\\                                        \label{ec2}
                         3{\beta'}^2 + e^{-2\beta} &=& \Lambda,
\\                                         \label{ec3}
                           - \beta'' + 3{\beta'}^2 &=& - \epsilon{\phi'}^2 - \Lambda.
\end{eqnarray}
    Equation (\ref{ec2}) (from which it follows $\Lambda > 0$) is solved giving
\begin{eqnarray}
                  e^\beta = \frac{\cosh(b u)}{\sqrt{3}b}, \qquad b := \sqrt{\Lambda/3}.
\end{eqnarray}
   Hence, the metric takes the following form:
\begin{eqnarray}
             ds^2 = \frac{9b^4}{\cosh^4(bu)}dt^2 - du^2 - \frac{\cosh^2 (bu)}{3b^2}d\Omega^2.
\end{eqnarray}

  With the expression found for $\beta$ and  \eqn{ec1} we find
\begin{eqnarray}
         \epsilon{\phi'}^2 = - 2b^2\biggr \{3 - \frac{2}{\cosh^2(bu)}\biggl\}.
\end{eqnarray}
  This solution exists only for $\epsilon = - 1$, that is, a phantom scalar field, for which
  we have the expression
\begin{eqnarray}
         \phi' = \pm b \sqrt{2}\sqrt{3 - \frac{2}{\cosh^2(bu)}},
\end{eqnarray}
  meaning that the scalar field is an always increasing or decreasing function of the spatial 
  coordinate $u$.

  The geometry of this solution exactly coincides with that of the second special solution found in 
  Ref. \cite{denis} for a $k$-essence field, described there in some detail (at $u = \pm \infty$ 
  there are two horizons of infinite area connected by a wormhole whose 
  throat is located on the sphere $u = 0$), but here we have again a different behavior of the 
  scalar field.

\subsection{Special solution III: $a = 3/2$}  

  This specific value of $a$ implies $\lambda \to \infty$ but again, with such a condition,
  the whole set of equations is well-behaved.

 If we put $ a = 3/2$, the field equations \eqref{EE-Ra} lead to $R=0$. (It should be stressed that 
 in this case the equations do not depend on the potential $V(\phi)$, which can be chosen
 arbitrarily at the outset.) 
 Together with the condition $R^0_0=R^2_2$, also following from the field equations and 
 the properties of the energy-momentum tensor, it gives
\bearr
         3R^0_0 + R^1_1 = 0 \ \ \then                                        \label{3-01}
\ \ 
	\beta'' + 2\gamma'' + (\beta'+ 2\gamma')(\beta'+ \gamma'-\alpha') = 0,   
\ear
  where we have used the general form of the metric \rf{ds} with an arbitrary radial 
  coordinate $u$ (not to be confused with the coordinate $u$ of the previous subsection), 
  and the prime again stands for $d/du$. Now, it is helpful to choose again the quasiglobal 
  gauge $\alpha + \gamma = 0$: we thus return to the notations $u=x$, $\e^\beta =r(x)$, 
  $\e^{2\gamma} = A(x)$, and the metric takes the form \rf{ds-q}.
  Then \eqn{3-01} takes an easily integrable form: 
\beq                                         \label{A*r}                                                      
         \beta'' + 2\gamma'' + (\beta'+ 2\gamma')^2 =0 \ \then\
         \e^{\beta + 2\gamma} \equiv  Ar = px + q,   
\eeq   
   where $p$ and $q$ are \intco s. On the other hand, the scalar field equation, as before,
   leads to the relation \rf{int-phi}, which for $a = 3/2$ reads
\beq                                         \label{phi3}
            \phi' = C A^{-5/6} r^{-4/3}. 
\eeq

   Now, with \rf{A*r} we have two qualitatively different situations, to be considered separately.

\medskip\noi
 {\bf 1.} $p\ne 0$. In this case, without loss of generality, we can put $q = 0$ by choosing the 
  zero point of $x$ and $p=1$ by choosing the scale of the time coordinate $t$. As a result, we have
\beq                                                      \label{A3a}
             A = x/r(x), \qquad   \phi' = C x^{-5/6} r^{-1/2}.       
\eeq
  It remains to find the function $r(x)$, which can be done using the first-order equation 
  \rf{eb2} with all necessary substitutions. We obtain 
\beq                   \label{r'3a}
		r' -1 = \frac{3\ep C^2}{4}\, x^{-2/3}, 
\eeq   
  whence
\beq                         \label{r3a}
                   r(x) = x + B\,x^{1/3} + r_0, \qquad        B: =\frac 94 \epsilon C^2, 
\eeq
  where $r_0$ is one more \intco. This completes the solution.

  It is clear that at $x =0$ the analyticity of this solution is violated, so it cannot be extended beyond
  this value, and it makes sense to restrict ourselves to $x > 0$. In any case, as follows from 
  \rf{A3a} and \rf{r3a}, as $x\to \infty$ we have a spatial infinity where $r \approx x$ and 
  $A \to 1$, but it is not a regular flat infinity because, instead of a Schwarzschild asymptotic, 
  $A(x)$ behaves at large $x$ as $A(x) \approx 1 - B x^{-2/3}$.

  The lower bound of the $x$ range depends on the parameters $B$ and $r_0$. If the function 
  \rf{r3a} has a zero at some $x_1 >0$, this zero corresponds to a repulsive central singularity 
  where $r\to 0$ and $A \to \infty$. This is possible if $B <0$, that is, if $\ep = -1$.

  If $r > 0$ at all $x \geq 0$, which is possible with sufficiently large $r_0$ even if $B < 0$, then 
  the solution is defined in the whole range $x \in \R_+$, and at $x=0$ we observe what seems to be
  a horizon, but there is no extension beyond it due to the term with $x^{1/3}$ that violates analyticity.

  If  $B>0$ and $r_0=0$, we have $r =0$ at $x=0$ and $r>0$ at $x>0$, the solution is again 
  defined for $x\in \R_+$, but now at $x=0$ we have an attractive singular center, $r=0$ and $A=0$.
  One can also find cases where the solution is defined between two singularities, one horizon-like 
  at $x=0$ and another at some $x > 0$ having the nature of a repulsive center. 

\medskip\noi
{\bf 2.}  $p=0$, in which case we have, instead of \rf{A3a},
\beq                               \label{A3b}
                        A(x) = q/r(x), \qquad        \phi' = C q^{-5/6} r^{-1/2}, \qquad q > 0.
\eeq           
  In this case, an attempt to find $r(x)$ from the first-order equation \rf{eb2} only leads to a 
  relation between the constants
\beq
			4 q^{2/3} = 3 C^2 
\eeq
  and to the restriction $\ep =-1$, that is, such a solution exists only for a phantom scalar field; 
  meanwhile, $r(x)$ remains arbitrary. However, this function can be found (equivalently) 
  from \eqn{0-1} or \rf{0-2} as follows:
\beq                                        \label{r3b}
                      r(x) = \frac{x^2}{3q} - Kx + r_0,           
\eeq                  
  with \intco s $K$ and $r_0$. The solution is defined in a range of $x$ where $r > 0$, depending 
  on these parameter values. Where $r=0$, we have a repulsive singular center, while in any of
  the limits $x \to \pm \infty$ we have a spatial infinity where $A\to 0$ (that is, a test mass is 
  attracted there), and the spatial geometry resembles that of a global monopole since there is an 
  incorrect limit of the circumference to radius ratio (we have $Ar'^2 \to 4/3$ instead of unity at usual 
  flat geometry).

  If the parameters satisfy the inequality $3q K^2 < 4r_0$, then $ r> 0$ in the whole range $x \in \R$, 
  and the solution describes a traversable wormhole that connects two such infinities.

\section{Regular solutions with a potential}

  In Ref. \cite{prl}, a self-interacting phantom scalar field in GR was considered, and globally 
  regular solutions were obtained by using the inverse problem method: a convenient 
  form for the radius $r(x)$ was chosen as a function of the quasiglobal coordinate $x$, for the
  metric written in the form 
\begin{eqnarray}                                                     \label {ds-qua}
                           ds^2 = A(x)dt^2 - A(x)^{-1}dx^2 - r(x)^2d\Omega,
\end{eqnarray}
  After that, the other metric function, $A(x)$, as well as the scalar field and its potential were 
  determined from the field equations, which coincide with \rf{eb1}--\rf{eb4} under the condition 
  $a=1$. 

  It is not hard to notice that \eqs  \rf{0-1} and \rf{0-2}, which are independent of both $a$ and $V$,
  are sufficient for finding the function $A(x)$ and $\phi(x)$ provided $r(x)$ is known. And even 
  more than that, these functions form a solution to the set of equation \rf{eb1}--\rf{eb4} of
  Rastall's theory for any value of $a$ except $a=3/2$. The only difference from GR will be in 
  the form of the potential $V(\phi)$ which will be now $a$-dependent.  

  Let us for convenience repeat here \eqs  \rf{0-1} and \rf{0-2}:
\bearr                        \label{0-1a}
                    2\frac{r''}{r} = - \epsilon{\phi'}^2,
\yyy                           \label{0-2a}
		  A'' r^2 - A (r^2)'' = -2
\ear
  The radial function $r(\rho)$ was fixed in \cite{prl} as
\begin{eqnarray}                                 \label{r(x)}
                        r(x) = \sqrt{x^2 + b^2},
\end{eqnarray}
  where $b$ is a constant with the dimension of length. Since with this choice  $r'' >0$, it follows
  from \rf{0-1a} that the scalar field is phantom, which provided a variety of solutions including 
  wormholes with flat or AdS asymptotics as well as regular black holes of the type that was 
  later named {\it black universes} \cite{bu-06}. The latter 
  are \asflat\ black holes in which an observer after crossing the horizon gets into an expanding 
  anisotropic universe with a Kantowski-Sachs type metric, which eventually tends to isotropy 
  and is asymptotically de Sitter at late times.

  This picture is completely reproduced in the model under study in the present work, with the  
  same expressions for $A(x)$ and $\phi(x)$ as in \cite{prl}. Thus, from \rf{r(x)} and \rf{0-1a}
  it follows
\begin{eqnarray}
                     \phi = \pm \sqrt{2}\arctan\frac{x}{b} + \phi_0,   \qquad \phi_0 = \const, 
\end{eqnarray}
  while for the metric function $A(x)$ we obtain from \rf{0-2a}
\begin{eqnarray}
           A = f_0 r^2 + 1 + \frac{x_0}{b^3}\biggr\{bx + r^2\arctan\frac{x}{b}\biggl\},
\end{eqnarray}
  where $f_0$ and $x_0$ are constants. Thus we have here the same geometric structures 
  as in \cite{prl}.

  The main difference between that case and ours resides in the expression for the potential which 
  now reads 
\begin{eqnarray}
              V(x) = - \frac{1}{(3 - 2a)r^4}\biggr\{r^2(A'x - 1) + A(x^2 + ab^2)\biggl\}
\end{eqnarray}
  which reduces to the case of GR when $a = 1$. This difference, maybe seeming tiny, may deeply  
  affect the stability of the model since $a$ is in general completely arbitrary in sign and 
  absolute value. The only exception is $a = 3/2$, in which case the above solution does not exist,
  and the system behaves as described in Subsection 4.3.  

  Since the energy-momentum tensor with zero trace is added to the Rastall equations 
  in the same way as to the Einstein equations, it can also be predicted that solutions with an \elmag\
  field \cite{we-cqg12} which generalize those obtained in \cite{prl} and are rich in various global 
  structures will also hold as solutions of Rastall's theory but only with modified scalar field potentials.  

\section{Conclusions}

  Rastall's theory \cite{rastall} is a nonconservative extension of GR in which the energy-momentum
  tensor of matter has a divergence proportional to the gradient of the Ricci scalar $R$. A simple 
  and immediate observation is that all solutions of GR in which $R\equiv 0$ are simultaneously 
  solutions of Rastall's theory with an arbitrary value of its free parameter $a$. 
This concerns not only the Schwarzschild and Reissner-Nordstr\"om solutions and their extensions 
  containing a massless conformally coupled scalar field \cite{k-73,boch,bek}; it is also appropriate
  to mention 
  a number of \ssph\ solutions with $R\equiv 0$ describing \bhs\ (including regular ones)
  \cite{bw-bh} and wormholes \cite{bw-wh} in quite a different extension of GR, the brane world 
  concept, in which Einstein-like gravity exists on a 4D subspace (brane) of a higher-dimensional 
  manifold (bulk)  \cite{RS2, SMS}. Thus all those wormhole and \bh\ configurations with zero $R$ 
  satisfy the Rastall equations with the same components of the energy-momentum tensor. 
  (For a more general study of wormhole properties in Rastall's theory see the recent paper \cite{R-wh}.)

  In this paper we studied the situations of greater interest from the viewpoint of Rastall's 
  theory,  considering physical systems where the free parameter $a$ plays a significant role. 
  We have used a canonical expression for the energy-momentum tensor of a scalar field, 
  in which case Rastall's theory takes a form similar to some class of Galileon 
  theories \cite{horn}. Exact solutions with a massless scalar field were found by fixing the free
  parameter $a$ (at $a = 1$, GR is recovered) at some particular values. Surprisingly, some of 
  these exact solutions are, from the geometric point of view, identical to those found in the context 
  of $k$-essence theories \cite{denis}. Such solutions have very particular features, some of them 
  being similar to the cold black hole case (e.g., they possess an infinite horizon area), but having
  singularities in the asymptotic region instead of flat space-time. We have also shown that, in the 
  absence of a potential term, it is in principle possible to have black holes with a finite horizon area 
  and with a finite and regular scalar field, contrary to what occurs in the $k$-essence case.
  In addition to these cases with similar features to those found in the $k$-essence context, 
  new structures were found, somewhat similar to the Schwarzschild solution of GR, but 
  without asymptotic flatness. 

  Moreover, it has been shown quite generally that it is possible, when the scalar field is massless, to 
  have structures with a regular horizon, admitting an analytical extension across the horizon, for 
  discrete values of $a$ in the interval $- 1 < a < 0$, but which are not asymptotically flat. 

  If a self-interaction potential term of the scalar field $V(\phi)$ is included, it is possible to obtain 
  regular solutions, completely equivalent to those existing in GR, the only difference being 
  in the expression for $V(\phi)$ which now depends on the Rastall parameter $a$. In particular, this
  concerns the regular wormhole and black hole (black universe) solutions found in Ref. \cite{prl} for 
  a scalar field of phantom nature. As in the previous case, this difference may substantially affect 
  the stability properties of the solutions.

  The stability of the structures found here and in the related references is a crucial subject, 
  and we intend to study it in our future work.

\subsubsection*{Acknowledgments}

  We thank FAPES (Brazil), CAPES (Brazil) and CNPq (Brazil) for partial financial support.
  The work of KB was performed within the framework of the Center 
  FRPP supported by MEPhI Academic Excellence Project 
  (contract No. 02.a03. 21.0005, 27.08.2013),
  and within the RUDN-University program 5-100.

\end{document}